\documentclass[a4paper, 12pt]{article}               
\usepackage[english]{babel}                          
\usepackage[T2A]{fontenc}                             
\usepackage[utf8]{inputenc}

\usepackage{mathtext}                                 
\usepackage{indentfirst}                              
\usepackage[unicode]{hyperref}                       
\usepackage{graphicx}
\usepackage{array,longtable,lscape}
\usepackage{amsmath,amsfonts,amssymb,amsthm,mathtools}
\usepackage{icomma}	                              
\usepackage{latexsym}
\usepackage{esint}
\usepackage{csquotes}

\usepackage{extsizes}

\usepackage[left=3cm, right=1.5cm, top=2cm, bottom=2cm]{geometry}

\usepackage[backend=biber, bibencoding=utf8, style=gost-numeric, sorting=none]{biblatex}
\title{Eliminating the Hubble Tension in the Presence of the Interconnection between Dark Energy and Matter in the Modern Universe}
\author{G. S. Bisnovatyi-Kogan \footnote{Space Research Institute, Russian Academy of Sciences, Moscow, Russia; Moscow Engineering Physics Institute (MEPhI), Moscow, Russia}, A. M. Nikishin\footnote{Moscow Engineering Physics Institute (MEPhI), Moscow, Russia}}
\date{}

\begin{document}
\maketitle

\begin{center}
    \Large \textbf{Abstract}
\end{center}

It is accepted in modern cosmology that the scalar field responsible for the inflationary stage of the early Universe is completely transformed into matter. It is assumed that the accelerated expansion is currently driven by dark energy (DE), which is likely determined by Einstein’s cosmological constant, unrelated to the scalar field responsible for inflation. We consider a cosmological model in which DE can currently have two components, one of which is Einstein’s constant ($\Lambda$) and the other, smaller dark energy variable component DEV ($\Lambda_V$), is associated with the remnant of the scalar field that caused inflation after the main part of the scalar field has turned into matter. We consider only the stages of evolution of the Universe after recombination ($z\lesssim 1100$), where dark matter (DM) is the predominant component of matter. It is assumed that the transformation of the scalar field into matter continues at the present time and is accompanied by the reverse process of the transformation of DM into a scalar field. The interconnection between DM and DEV, which leads to a linear relationship between the energy densities of these components after recombination $\rho_{DM}=\alpha\;\rho_{DEV}$, is considered. Variants with a dependence of the coefficient $\alpha(z)$ on the redshift $z$ are also considered. One of the problems that have arisen in modern cosmology, called Hubble Tension (HT), is the discrepancy between the present values of the Hubble constant ($H_0$) measured from observations of the Universe at small redshifts ($z\lesssim 1$) and the values found from fluctuations of the cosmic microwave background in the Universe at large redshifts ($z\approx 1100$). In the model under consideration, this discrepancy can be explained by the deviation of the existing cosmological model from the conventional cold dark matter (CDM) model of the flat Universe by the action of the additional dark energy component DEV at the stages after recombination. Within this extended model, we consider various $\alpha(z)$ functions that can eliminate the HT. To maintain the ratio of DEV and DM energy densities close to constant over the interval $0\leq z \lesssim 1100$, it is necessary to assume the existence of a wide spectrum of dark matter particle masses.

\newpage
\section{Introduction}

Hubble’s law $v=Hr$ is one of the most important laws of cosmology, which sets the recession velocity of space objects $v$ depending on the distance between them $r$. The measurement of parameter $H_0$ in the present era is a very difficult and non-trivial task, which has been undertaken by various scientific groups for many years. The recession velocity is determined from the redshift of the lines observed in the spectra of receding galaxies, but the greatest difficulty lies in measuring the distances to them. For this purpose, the cosmic distance ladder method is applied, which uses increasingly brighter standard candles, i.e., objects with a known luminosity. This method allows measuring distances to relatively close objects, and is not free from systematic errors associated with the inevitable spread of the luminosities of the standard candles. The use of different steps in this ladder by different groups led in 1972– 1974 to a significant discrepancy between the $H_0$ values: from $\sim\; 50 $ km/s/Mpc by the Sandage–Tammann group to $\sim \; 100 $ km/s/Mpc by the Vaucouleur group \cite{ZeldNov}. Over time, $H_0$ measurements at redshifts $\lesssim 1$ have been greatly improved by the development of large telescopes, including measurements on the Hubble telescope, which have made possible to limit the range of values to $H_0 \approx 72-75 $ km/s/Mpc.

Measurements of the CMB fluctuation spectrum on the WMAP and Planck satellites led to the possibility of independent measurement of the Hubble parameter $H_r$ during the recombination epoch. Using the theoretical $H(z)$ dependence within the accepted cosmological model, presumably $\Lambda CDM$, the modern value of the Hubble parameter $H_0^{Dist}$ was obtained, which differed from $H_0^{Loc}$ found from local measurements by a statistically significant value within $4.5\sigma \; - \; 6.3\sigma$ \cite{Riess2}. This discrepancy is the essence of the HT problem (see, however, \cite{Friedmann3}).

Currently accepted cosmological models assume that the scalar field responsible for the existence of the early stages of rapid exponential expansion completely turns into matter during inflation, and the modern accelerated recession of galaxies is caused by dark energy (DE), presumably Einstein’s constant $\Lambda$, which is unrelated to the scalar field that led to the inflation.

According to the measurements of the relict fluctuations (WMAP, Planck), our Universe is flat with an accuracy $< 1\%$, and its average density is equal to the critical density $\rho_c$. The modern $\Lambda CDM$ model of the Universe is characterized by the following parameters:

\begin{equation}
\label{univ}
    \begin{aligned}
        &\text{contribution of non-relativistic matter} \quad  &&\Omega_M=\Omega_{DM}+\Omega_B \approx  0.26+0,04 \approx 0.3.   \\
        &\text{dark energy contribution} \quad  &&\Omega_{\Lambda} \approx 0.7. 
    \end{aligned}
\end{equation} 
Here, $\Omega_i = \rho_{i,0}/\rho_{c}$; $\Omega_{\Lambda}$ is determined by the Einstein constant $\Lambda$; $\Omega_{DM}$ and $\Omega_B$ are determined by the density of dark matter and baryonic matter.

In this paper, following \cite{BisnKog_arxiv,BisnKog_Universe}, in order to explain the HT discrepancy, we consider a cosmological model that is an extension of the conventional $\Lambda CDM$ model and includes two components of DE, one of which is the Einstein constant ($\Lambda$), and the other, a small variable component DEV ($\Lambda_V$), is the remainder of the scalar field that created the inflation. The appearance of the HT problem is associated with the recalculation of the Hubble parameter $H^{Dist}(z_r)$ measured at the time of recombination to the modern $H^{Dist}(0)$ value using the standard $\Lambda CDM$ model, which is assumed to be incomplete. Therefore recalculation leads to an underestimated value of the modern Hubble constant. In this study, a more correct solution of the cosmological equation is obtained, which considers the action of the additional variable component of DEV, which significantly refines the results obtained in \cite{BisnKog_arxiv,BisnKog_Universe} for DEV density values at which the HT problem does not arise.

\section{Hubble Tension}
A problem that has arisen in cosmology in recent years is the discrepancy in the values of the Hubble constant at the present epoch obtained in different experiments. The analysis of the observations of the Planck mission, which measures the CMB fluctuations during the recombination period, leads to the current value of the Hubble constant \cite{WMAP,Planck2015,Planck2018}:
\begin{equation}
    H_{0}^{Dist} = 67.36 \pm 0.54 \; {\mbox{ km/s/Mpc}}.
\end{equation}
At the same time, measurements that involve type Ia supernovae (SNIa) with distance calibration using Cepheids \cite{Riess14,Perlmutter15,Riess16,Riess17,Riess18} give the value
\begin{equation}
    H_{0}^{Loc} = 74.03 \pm 1.42 \;{\mbox{ km/s/Mpc}}.
\end{equation}
Measurements using the time delays of lensed quasars \cite{Wong19} give $H_0 = 73.3_{-1.8}^{+1.7}\;$ km/s/Mpc. The value $H_0 = 72.4 \pm 1.9 \;$ km/s/Mpc was found in \cite{Yuan20} using the red giant branch applied to SNIa, which is independent of the Cepheid distance scale. The analysis of a set of these and other recent measurements at small and large redshifts shows \cite{Verde21} that the discrepancy
between Planck’s results \cite{Planck2018} and any three independent measurements in the late Universe lies between $4\sigma$  and $6\sigma$. Several new space experiments have been proposed to test the reliability of this discrepancy in the values of the Hubble constant \cite{Bengaly32,Valentino33}.

Many different explanations for the occurrence of the HT have been proposed, some of which have been refuted by observational evidence. There is currently no generally accepted and experimentally proven solution to the HT problem. The proposed solutions can be divided into pre-recombination and post-recombination solutions, which in some way change the process of evolution of the Universe either in the period before recombination or after it, respectively. Variants with the use of gravity theories based on modifications of general relativity are also considered (see \cite{Karwal22,Moertsell23,Poulin24,Yang25,Vagnozzi26,Valentino27,Umilta28,Ballardini29,Rossi30,Knox31,Lucovic157,Kenworthy156,Sakstein205,Gogoi206,Zhao264,Mortonson316,Li391,Parker269,Steigman403,Amendola554,Lin188,Hu796}). A detailed review of most of the proposed methods for solving this problem is given in \cite{Valentino33}. It outlines proposed ways to explain this phenomenon, as well as the possible impact of their presence on other cosmological parameters.

There is a significant discrepancy between the experimental $H_0$ values obtained in the Planck mission from the CMB fluctuations and the values obtained in local measurements. In this regard, we will consider the possibility of solving the HT problem as a discrepancy between the results of these experiments considering the average value obtained in local measurements as true. In further numerical calculations, we use the $H_0^{Dist}$ value obtained as a result of measurements at high redshifts and the $H_0^{Loc}$ value obtained as a result of measurements in the local Universe.

\section{Suggested solution to the HT problem}

Dark matter and DE make up about $96\%$ of all energy in the Universe \cite{WMAP,Riess14,Perlmutter15}, but their nature is still unknown. The current value of the DE density can be represented by Einstein’s cosmological constant $\Lambda$ \cite{Einstein34}, but it can also be the result of the Higgs scalar field, which is proposed as the cause of inflation in the early Universe \cite{Guth,Linde,Starobinsky,Mukhanov}. The value of induced $\Lambda_V$ related to inflation is many orders of magnitude greater than its current value, and no attempts have been made to find a connection between them. The physical nature of DM remains unclear. There are many assumptions about its origin \cite{Arun,Zhao,Samart}, but none of them has been confirmed experimentally or observationally, while many of them have been refuted.

To explain the origin of the HT, the variable part of dark energy (DEV) was introduced in \cite{BisnKog_Universe,BisnKog_arxiv} as a variable component of the cosmological constant $\Lambda_{V}$. Its contribution to the process of expansion of the Universe increases with $z$ and exceeds the contribution of the Einstein’s cosmological constant $\Lambda$ at redshifts $z\gg 1$. Thus, DE can have two components, one of which is Einstein’s constant $\Lambda$, and the other, a currently small variable component DEV, $\Lambda_V$, which can originate from the remnants of the inflationary scalar field, which is the origin of existing matter. It was assumed that the process of transition of field energy into matter is currently accompanied by the reverse process of transformation of mass into field energy, which leads to a dynamic relationship between the matter densities and $\Lambda_V$.
 
We consider only the post-recombination period ($z \lesssim 1100$) of the expansion of the universe, where DM is the most significant component of matter in the universe. Thus, for simplicity, we assume that there is a relationship between the densities of the dark matter energy and the additional part of DE:
\begin{equation}
\label{din}
    \rho_{DM} = \alpha(z) \rho_{DEV}.
\end{equation}
Here, $\alpha(z)$ is a function of redshift, the form of which is constrained by the requirement that it allows the HT contradiction to be eliminated without creating additional difficulties in interpreting cosmological observations. As shown in \cite{BisnKog_Universe,BisnKog_arxiv}, this requirement is satisfied, for example, by a constant $\alpha$ value that eliminates the HT. To maintain dynamic equilibrium (\ref{din}), dark matter must consist of a spectrum of particles of various masses, including very light particles. The birth of such particles in the process of mutual transformation of DM and scalar field over the entire interval $z \lesssim 1100$ will allow maintaining a dynamic equilibrium between the DM and DEV densities of type (\ref{din}) with a decrease in the energy of all variable components of the Universe during its expansion.
  
As follows from the numerical simulation of the processes that lead to the formation of the modern large-scale structure of the Universe, the model with cold DM shows the best agreement with observations, i.e., DM particles are nonrelativistic. Massive particles from supersymmetric field theory (neutralino, photino, etc. \cite{RubGorb}) are often considered for this. On the other hand, very light axion particles \cite{Arun} are considered as DM; their existence follows from some theoretical models that explain the observed violation of CP invariance in nuclear processes. In our model of the Universe, eliminating the HT requires a more complex DM structure, where particles of intermediate masses can be present, and the mass of light particles (axions) can also be represented by a whole spectrum.

In the presence of DEV, the Hubble constant decreases with time more slowly than without it. This creates a greater contemporary $H_0$ value for the same $H_{rec}$ value in the recombination era. Therefore, we assume that the $H_{0}^{Dist}$ value measured by the Planck mission was obtained by extrapolating the $H_{rec}$ value from the time of recombination $z_{r} \approx 1100$ to the present $z = 0$ in the Friedmann model of a flat dusty Universe, considering the cosmological constant $\Lambda$. In the case of some sort of equipartition in the Universe in the form of (\ref{din}), this extrapolation should be carried out in a model that considers the additional components of the dark energy DEV. In our interpretation, the HT is related to an inaccurate extrapolation of the Planck data within a model that does not consider DEV.

Let us assume that the true modern value of the Hubble constant is determined by local measurements, that is $H_{0}^{Loc}$, and both measurements are correct, but the $H_{0}^{Dist}$ value occurred due to inaccurate extrapolation. Knowing this, it is possible to find such a function $\alpha(z)$ that the recalculation of the Planck data yielded the value of the Hubble constant that coincided with local measurements. Thus, the problem is reduced to finding such a function $\alpha(z)$ for which the HT does not currently arise in the process of recalculating the Hubble constant. In order to calculate the effect of a small DEV addition on the calculation of the modern $H_0$ value from the $H_{rec}$ measurements at the time of recombination, we must construct a cosmological model considering the DEV component on the interval $z \in [0, z_r]$.

\section{Expansion of the Universe}

The evolution of the Universe is described by Einstein’s GR equations \cite{ZeldNov}:
\begin{equation}
\label{gre}
    R_{ik} - \frac{1}{2}g_{ik} R - \Lambda g_{ik} = \frac{8\pi G}{c^4} T_{ik}.
\end{equation}
As shown by A.A. Friedmann in 1922 \cite{Friedmann}, the expansion of a homogeneous isotropic Universe is determined by one equation for the scale factor, $a(t)$, which follows from the (00) component of the Einstein equation (\ref{gre}), in the form
\begin{gather}
\label{eq}
    \left( \frac{\dot{a}}{a} \right)^2 = \frac{8\pi G}{3} \rho  + \frac{\Lambda c^2}{3} -\frac{\kappa c^2}{a^2},
\end{gather}
where $\rho\, c^2$ is the total energy density in the Universe, the second term on the right in (\ref{eq}) is related to the cosmological constant, and the last term determines the contribution of the scalar space curvature. The best agreement between theory and observations takes place for a flat universe with $\kappa = 0$, which is considered below. To relate pressure and density, the adiabatic expansion condition is used

\begin{equation}
\label{ad}
    \frac{d\rho}{\rho+P} = -\frac{d\mathcal{V}}{\mathcal{V}} = -3\frac{da}{a}, \quad \text{where $\mathcal{V} - $ is volume.}
\end{equation}
Wherein
$$ \rho + P = \rho_{DM} + \rho_{DEV} + P_{DM} + P_{DEV} \equiv \rho_m+\rho_\phi+P_m+P_\phi. $$

\section{Scalar field as the source of DM and DEV}

The scalar field is accepted as the main cause of the inflationary stage and the creation of matter in the Universe \cite{Guth,Linde,Starobinsky}. Let us consider a scalar field of intensity $\phi$, which is in the potential $V(\phi)$. In a homogeneous isotropic expanding Universe, the time dependence of $\phi$ is described by the equation \cite{Peebles}
$$ \ddot{\phi} + 3\frac{\dot{a}}{a}\dot{\phi} = -\frac{dV}{d\phi}. $$
The energy density $\rho_V$ and pressure $P_{V}$ of the scalar field are defined as \cite{Peebles} (here and in most of the subsequent equations, $c=1$)

\begin{equation}
\label{ep}
 \rho_V =  \frac{\dot{\phi}^{2}}{2} + V(\phi), \quad \quad   P_V = \frac{\dot{\phi}^{2}}{2} - V(\phi). 
\end{equation}

Let us consider the Universe with initial scalar field intensity $\phi_{in}$, initial potential $V_{in}$, and zero initial derivative $\dot\phi_{in}=0$. The derivative of the intensity of the scalar field increases in the initial stages of inflation as the potential $V$ decreases. Let us assume that after achieving the validity of the relation

\begin{equation}
\label{eqp}
   \dot{\phi}^{2} = 2\alpha(z) V
\end{equation}
it continues to hold true in further expansion stages. The kinetic part of the energy of the scalar field turns into matter, presumably mainly into dark matter, and the $\alpha(z)$ function determines the relationship between the density of dark energy (DE), determined by the $V$ value, and the density of matter, determined by the kinetic term. As follows from observations, the bulk of DE can currently be related to Einstein’s constant $\Lambda$. At previous stages of expansion, the constant $\Lambda$ was smaller than the variable part $\Lambda_V$ in a wide range of admissible functions $\alpha(z)$. If the condition (\ref{eqp}) is satisfied, we introduce the following notation

\begin{equation}
\label{eqp1}
 \rho_\phi=V,\quad P_\phi=-V,\quad  \rho_m=\frac{\dot\phi^2}{2},\quad P_m=\beta\frac{\dot\phi^2}{2}.
\end{equation}
for $\rho_m=\alpha(z)\rho_{\phi}$, $0\leq \beta <1$. It is assumed here that the pressure of the substance formed from the dynamic component of the field energy is lower than the field pressure, because when the massless phase passes into particles with a nonzero rest mass, part of the energy passes into the rest mass, and the kinetic energy that creates pressure decreases down to zero, $0\leq \beta \leq1$. During the expansion of the Universe, a continuous phase transition occurs between state (\ref{ep}), which corresponds to the free field, and state (\ref{eqp1}), where the kinetic part of the field has turned into matter. Using (\ref{eqp}) and (\ref{eqp1}), we obtain

\begin{equation}
\label{eqp2}
 \rho=\rho_\phi+\rho_m=(1+\alpha)\,V,\quad 
 P= P_\phi+P_m=-(1-\alpha\beta)\,V.
\end{equation}
Note that in this case, the dynamic equilibrium between the field and matter is maintained in the form (\ref{eqp}), (\ref{eqp2}). Solving the adiabatic equation (\ref{ad}) for constants $\alpha$ and $\beta$, and considering (\ref{eqp2}), we obtain \cite{BisnKog_Universe,BisnKog_arxiv}

\begin{equation}
\label{ad1}
 \dot\rho=-3\alpha\frac{1+\beta}{1+\alpha}\frac{\dot a}{a}\,\rho, \quad \frac{\rho}{\rho_\star}=\left(\frac{a_\star}{a}\right)^{\frac{3\alpha(1+\beta)}{1+\alpha}}.   
\end{equation}
Here, the asterisk denotes the known values of the functions at an arbitrarily chosen time point $t_\star$ corresponding to the redshift $z_\star$. Further considering the Universe at the stage after recombination, we will use the dust matter approximation with $\beta=0$.

\section{Cosmological models and solution of the HT problem}

The solution of cosmological equations (\ref{eq}) and (\ref{ad1}) for a flat dusty Universe after recombination at $\kappa=\beta=0$, and constant $\alpha$ was considered in \cite{BisnKog_arxiv,BisnKog_Universe}. We present solutions for two cases.

\subsection { \texorpdfstring{ $\Lambda = 0$,    $\alpha(z) = const$}{}} 

In this case, we obtain the following time dependences of the functions:

\begin{gather}
\label{anul}
    \frac{a}{a_*} = \left( 6\pi G\rho_* t^2 \right)^{\frac{1+\alpha}{3\alpha}} \left[ \frac{\alpha}{1+\alpha} \right]^{\frac{2(1+\alpha)}{3\alpha}} = \left( \frac{\rho_*}{\rho}\right)^{\frac{1+\alpha}{3\alpha}} = \left(\frac{t}{t_*} \right)^{\frac{2(1+\alpha)}{3\alpha}}, \\
    \rho = \left( \frac{1+\alpha}{\alpha}\right)^2 \frac{1}{6\pi G t^2}
    \label{anul1},
\end{gather}
Choosing the modern age of the Universe $t_0$ as an arbitrary value $t_\star$ and using the relationship between the redshift and the scale factor, we obtain
\begin{gather*}
    z+1 \equiv \frac{a_0}{a} = (6\pi G \rho_0 t^2)^{-\frac{1+\alpha}{3\alpha}} \left[\frac{\alpha}{1+\alpha} \right]^{-\frac{2(1+\alpha)}{3\alpha}} = \left( \frac{t_0}{t}\right)^{\frac{2(1+\alpha)}{3\alpha}}, \\
    t = \frac{1+\alpha}{\alpha \sqrt{6\pi G\rho_0}} (z+1)^{-\frac{3\alpha}{2(1+\alpha)}}.
\end{gather*}
Further, everywhere the subscript $\alpha$ will mean that this value is calculated in the model considering DEV. Using the definition of the Hubble constant and relation (\ref{anul}), we obtain
\begin{equation}
\label{hubn}
    H_{\alpha} \equiv \frac{\dot{a}}{a} = \frac{2(1+\alpha)}{3\alpha t} = \frac{2}{3} \sqrt{6\pi G\rho_0} (z+1)^{\frac{3\alpha}{2(1+\alpha)}}.
\end{equation}
For the limiting case $\alpha \to \infty$, in the absence of the DEV contribution, we have

\begin{gather*}
    \frac{a}{a_0} = \left(  6\pi G\rho_0 t^2 \right)^{\frac{1}{3}} = \left(\frac{\rho_0}{\rho}\right)^{\frac{1}{3}} = \left(\frac{t}{t_0}\right)^{\frac{2}{3}}, \quad \rho = \frac{1}{6\pi G t^2}, \quad H \equiv \frac{\dot{a}}{a} = \frac{2}{3t}, \\
    z+1 \equiv \frac{a_0}{a} = \left(6\pi G\rho_0 t^2\right)^{-\frac{1}{3}} = \left(\frac{t_0}{t}\right)^{\frac{2}{3}}, \,\,\, t = \frac{1}{\sqrt{6\pi G\rho_0}} (z+1)^{-\frac{2}{3}},\,\,\,
 H = \frac{2}{3}\sqrt{6\pi G\rho_0} (z+1)^{\frac{3}{2}}.
\end{gather*}
Here, we used solution of (\ref{anul1}) for $\rho(t)$ with zero arbitrary integration constant. A nonzero integration constant greatly complicates the process of finding an analytical solution. Instead, within this solution method, we slightly modify the obtained formulas for the Hubble parameter. In order to ensure the fulfillment of the boundary condition $H = H_{\alpha}$ at the time of recombination, we add an additional factor to expression (\ref{hubn}) and obtain
\begin{gather}
\label{hubn1}
    H_{\alpha}(z) = \frac{2}{3} \sqrt{6\pi G\rho_0}\; (z_r + 1)^{\frac{3}{2(1+\alpha)}} (z+1)^{\frac{3\alpha}{2(1+\alpha)}} \\
    \label{hubn1a}
    H(z) = \frac{2}{3} \sqrt{6\pi G\rho_0} \;(z+1)^{\frac{3}{2}}.
\end{gather}
In this solution, the $H$ values are the same at the time of recombination, but due to the different law of the expansion of the Universe, they have different modern values. The value obtained without considering DEV is identified with $H_0^{Dist}$, which turns out to be less than the locally measured $H_0^{Loc}$. If we assume that DEV is present in the Universe, the current $H_\alpha$ value should coincide with $H_0^{Loc}$. Equating the difference $H_0^{Loc}-H_0^{Dist}$ to the modern difference $H_\alpha-H$ from (\ref{hubn1}) and (\ref{hubn1a}), we obtain the $\alpha$ value at which the HT does not occur. The equation that determines the $\alpha$ value and its solution for $\alpha$ at which this paradox does not arise have the following form.

\begin{gather}
\label{hubn2}
    \frac{H_{0}^{Loc}}{H_{0}^{Dist}} = (z_r + 1)^{\frac{3}{2(1+\alpha)}}, \qquad \alpha_{HT} \approx 133.
\end{gather}
The values of the cosmological parameters used here and further are given in Table 1.

\begin{table}[h!]
\caption{Averaged cosmological parameters}
\begin{tabular}[b!]{|p{0.6\linewidth}|p{0.35\linewidth}|}
\hline
Parameter  &  Value \\
\hline
Local value of the Hubble constant      &  $\; H_{0}^{Loc} \approx  73 \; km/s/Mpc$     \\
\smallskip  &  \smallskip \\
Hubble constant measured from the CMB    &  $\; H_{0}^{Dist} \approx  67.5 \; km/s/Mpc$\\
\smallskip  &  \smallskip \\
Total density of matter in a flat universe   & $\; \rho_{tot} \approx 1.066 \cdot 10^{-29}$  $g/cm^3$\\
\smallskip  &  \smallskip \\
Locally measured density of the cosmological constant \cite{Perlmutter15} (2$\sigma$ statistics)  &  $\; \rho_{\Lambda}$ = $(0.44 \;\div \; 0.96)\; \rho_{tot}$\\
\smallskip  &  \smallskip \\
Distantly measured density of the cosmological constant  &  $ \rho_{\Lambda} \approx 0.7 \rho_{tot}$\\
\smallskip  &  \smallskip \\
Cosmological constant from remote measurements    &  $\; \Lambda = \frac{8\pi G\rho_{\Lambda}}{c^2} \approx 1.40 \cdot 10^{-56}$  $cm^{-2}$  \\
\smallskip  &  \smallskip \\
Asymptotics of the Hubble constant    &  $H_{ac} = \sqrt{\frac{\Lambda c^2}{3}} \approx 63.2\, km/s/Mpc$  \\
\smallskip  &  \smallskip \\
Average age of the Universe \cite{age}    &  $\; t_0 \approx 4.35 \cdot 10^{17}\; s$\\
\smallskip  &  \smallskip \\
Redshift corresponding to the recombination epoch   &  $\; z_{r} \approx 1100$ \\
\hline 
\end{tabular}
\end{table}

\subsection{\texorpdfstring{$\Lambda \neq 0$,    $\alpha(z) = const$}{}} 

Modern observations (\ref{univ}) indicate the predominance of DE determined by the cosmological constant $\Lambda$. Considering the additional variable term DEV, the parameters of the dusty Universe after recombination ($z < 1100$) are described by formulas obtained from the solution of system (\ref{eq}), (\ref{ad1}) in the form \cite{BisnKog_arxiv,BisnKog_Universe}:
\begin{gather}
\label{hubn3}
    \left(\frac{a}{a_*}\right)^{\frac{3\alpha}{2(1+\alpha)}} = \sqrt{\frac{8\pi G\rho_*}{\Lambda c^2}} \; \sinh{\left( \sqrt{\frac{\Lambda}{3}}\frac{3\alpha}{2(1+\alpha)} ct \right)} = \sqrt{\frac{\rho_*}{\rho}}, \\
    \label{hubn4}
    \sqrt{\frac{\Lambda c^2}{8\pi G\rho}} = \sinh{\left(\sqrt{\frac{\Lambda}{3}}\frac{3\alpha}{2(1+\alpha)} ct \right)}.
\end{gather}
Identifying the time point $t_*$ with the present age of the Universe $t_0$, we find the relationship between redshift and time in the form
\begin{gather}
\label{hubn5}
    z+1 \equiv \frac{a_0}{a} = \left[ \sqrt{\frac{8\pi G\rho_0}{\Lambda c^2}} \; \sinh{\left( \sqrt{\frac{\Lambda}{3}}\frac{3\alpha}{2(1+\alpha)} ct \right)} \right]^{-\frac{2(1+\alpha)}{3\alpha}} = \left( \frac{\rho}{\rho_0} \right)^{\frac{1+\alpha}{3\alpha}}.
\end{gather}
For the Hubble parameter from (\ref{hubn3}), we have the following expression:
\begin{gather}
\label{hubn6}
    H_{\alpha}(t) \equiv \frac{\dot{a}}{a} = \sqrt{\frac{\Lambda c^2}{3}} \; \coth{\left( \sqrt{\frac{\Lambda}{3}} \frac{3\alpha}{2(1+\alpha)} ct\right)}.
\end{gather}
The flat Universe without DEV is described by the relations following from (\ref{hubn3}), (\ref{hubn4}) in the limit at $\alpha \to \infty$
\begin{gather}
\label{hubn7}
    \left(\frac{a}{a_0}\right)^{\frac{3}{2}} = \sqrt{\frac{8\pi G\rho_0}{\Lambda c^2}} \; \sinh{\left( \sqrt{\frac{\Lambda}{3}}\frac{3}{2} ct \right)} = \sqrt{\frac{\rho_0}{\rho}}, \qquad 
    \sqrt{\frac{\Lambda c^2}{8\pi G\rho}} = \sinh{\left(\sqrt{\frac{\Lambda}{3}}\frac{3}{2} ct \right)},\\
    \label{hubn8}
    z+1 \equiv \frac{a_0}{a} = \left[ \sqrt{\frac{8\pi G\rho_0}{\Lambda c^2}} \; \sinh{\left( \sqrt{\frac{\Lambda}{3}}\frac{3}{2} ct \right)} \right]^{-\frac{2}{3}} = \left( \frac{\rho}{\rho_0} \right)^{\frac{1}{3}},\quad H(t) = \sqrt{\frac{\Lambda c^2}{3}} \; \coth{\left( \sqrt{\frac{\Lambda}{3}} \frac{3}{2} ct\right)}.
\end{gather}
We express time $t$ in terms of the observed redshift $z$:
\begin{flalign*}
      t_\alpha = \sqrt{\frac{3}{\Lambda c^2}} \frac{2(1+\alpha)}{3\alpha} \; \sinh^{-1}{\left[ \sqrt{\frac{\Lambda c^2}{8\pi G\rho_0}} (z+1)^{-\frac{3\alpha}{2(1+\alpha)}} \right]},\\
     t = \sqrt{\frac{3}{\Lambda c^2}} \frac{2}{3} \; \sinh^{-1}{\left[ \sqrt{\frac{\Lambda c^2}{8\pi G\rho_0}} (z+1)^{-\frac{3}{2}} \right]}\qquad.
\end{flalign*}
The Hubble parameter as a function of redshift is written as
\begin{flalign}
\label{hubn9}
   H_{\alpha}(z) = \sqrt{\frac{\Lambda c^2}{3}} \; \coth{\left(\sinh^{-1}{\left[ \sqrt{\frac{\Lambda c^2}{8\pi G\rho_0}} (z+1)^{-\frac{3\alpha}{2(1+\alpha)}} \right]} \right)},& \\
   \label{hubn10}
   H(z) = \sqrt{\frac{\Lambda c^2}{3}} \; \coth{\left(\sinh^{-1}{\left[ \sqrt{\frac{\Lambda c^2}{8\pi G\rho_0}} (z+1)^{-\frac{3}{2}} \right]} \right)}.&
\end{flalign}
The contribution of $\Lambda$ during recombination is very small, so expressions (\ref{hubn9}) and (\ref{hubn10}) can be expanded into a Taylor series, leaving only the first term. At the time of recombination, we then obtain
\begin{flalign}
 \label{hubn11}
    H_{r\alpha} \approx \frac{2}{3} \sqrt{6\pi G\rho_0} (z_r+1)^{\frac{3\alpha}{2(1+\alpha)}},\qquad
  H_r \approx \frac{2}{3} \sqrt{6\pi G\rho_0} (z_r+1)^{\frac{3}{2}}.
\end{flalign}
Similarly to the case $\Lambda=0$, we modify the solutions in (\ref{hubn9}) and (\ref{hubn10}) so that the boundary condition $H_{r\alpha} = H_{r}$ is satisfied at the time of recombination. We then obtain
\begin{eqnarray}
\label{hubn12}
    H_{\alpha}(z) = \sqrt{\frac{\Lambda c^2}{3}} \; \coth{\left(\sinh^{-1}{\left[\sqrt{\frac{\Lambda c^2}{8\pi G\rho_0}}(z+1)^{-\frac{3\alpha}{2(1+\alpha)}}(z_r + 1)^{-\frac{3}{2(1+\alpha)}} \right]} \right)}, \\
    \label{hubn13}
    H(z) = \sqrt{\frac{\Lambda c^2}{3}} \; \coth{\left(\sinh^{-1}{\left[\sqrt{\frac{\Lambda c^2}{8\pi G\rho_0}}(z+1)^{-\frac{3}{2}}\right]} \right)}.
\end{eqnarray}
These solutions must comply with $H_{0\alpha}=H_{0}^{Loc}, \; H_0=H_{0}^{Dist}$ at $z=0$, which uniquely determines the  $\alpha$ value. From (\ref{hubn12}) and (\ref{hubn13}), we obtain:
\begin{gather}
    (z+1)^{-\frac{3\alpha}{2(1+\alpha)}} (z_r + 1)^{-\frac{3}{2(1+\alpha)}} \sqrt{\frac{\Lambda c^2}{8\pi G\rho_0}} = \sinh{\left[\coth^{-1}{\left(\frac{H_{\alpha}}{\sqrt{\Lambda c^2/3}}\right)} \right]}, \\
    (z+1)^{-\frac{3}{2}}  \sqrt{\frac{\Lambda c^2}{8\pi G\rho_0}} = \sinh{\left[\coth^{-1}{\left(\frac{H}{\sqrt{\Lambda c^2/3}}\right)} \right]}.
\end{gather}
Hence, for $z=0$, $z_r=1100$ we obtain an equation that determines the parameter $\alpha$ at which the HT problem is eliminated. This equation and its numerical solution have the form
\begin{equation}
\label{ht1}
    \frac{\sinh{\left[\coth^{-1}{\left(\frac{H_{0}^{Loc}}{\sqrt{\Lambda c^2/3}}\right)} \right]}}{\sinh{\left[\coth^{-1}{\left(\frac{H_{0}^{Dist}}{\sqrt{\Lambda c^2/3}}\right)} \right]}} = (z_r + 1)^{-\frac{3}{2(1+\alpha)}}, \qquad \alpha_{HT} \approx 24.
\end{equation}
Assuming the full contribution of matter $\Omega_m\approx 0.3$, we obtain the modern DEV contribution to the density of the universe $\Omega_{DEV}= \frac{0.3}{24}\approx 0.0125$. If the contribution of DE is $\Omega_{DE}\approx 0.7$, it can consist of two components, $\Omega_{DEV}\approx 0.0125$ and $\Omega_{\Lambda}\approx 0.6875$, with the contribution of baryonic matter $\Omega_b\approx 0.04$.

\section{Solution of the problem using the \texorpdfstring{$H(z)$}{} function}

If we limit the problem by excluding the question of constructing a cosmological model in the form of dependence $a(t)$ and finding only the behavior of the Hubble parameter in the form of dependence $H(z)$, the problem of solving the HT problem becomes much simpler. In particular, it is possible to solve it analytically for a whole set of functions $\alpha(z)$. Cosmological equation (\ref{eq}) and adiabatic equation (\ref{ad1}), at $\kappa=\beta=0$, can be written as

\begin{equation}
   \label{eq1}
  H^2 = \frac{8\pi G}{3} \rho  + \frac{\Lambda c^2}{3}, \qquad  
 \frac{\dot\rho}{\rho}=-\frac{3\alpha}{1+\alpha}\frac{\dot a}{a}.
\end{equation}
Let us first consider the case of constant $\alpha$ and nonzero $\Lambda$, which was solved in Section 6.2.

\subsection{\texorpdfstring{$\alpha = const, \qquad \Lambda \neq 0$}{}}

Similar to the previous method, we first consider the case of a constant function $\alpha(z)$. The second equation in (\ref{eq1}) for this case is integrated in (\ref{ad1}). We write this solution in the form

\begin{equation}
   \label{eq2}
\frac{\rho_\alpha}{\rho_{0\alpha}}=
\left(\frac{a_{0\alpha}}{a_\alpha}\right)^{\frac{3\alpha}{1+\alpha}}=(1+z)^{\frac{3\alpha}{1+\alpha}}.
\end{equation}
Here, the present time is chosen as an arbitrary reference point, $\rho_\star=\rho_{0\alpha}$, and the definition of redshift $z+1=\frac{a_{0\alpha}}{a_\alpha}$ is used. It is also considered that $z_\alpha \equiv z$. Substituting (\ref{eq2}) into (\ref{eq1}), we obtain the dependence for the Hubble parameter in the form

\begin{gather}
\label{eq3}
    H_{\alpha}(z) = \sqrt{ \frac{\Lambda c^2}{3} + \frac{8\pi G}{3}\rho_{0\alpha} (1+z)^{\frac{3\alpha}{1+\alpha}}}.
\end{gather}
In the limit $\alpha \to \infty$, we obtain the dependence of the Hubble constant in the model that does not consider the DEV contribution:
\begin{gather}
\label{eq4}
    H(z) = \sqrt{\frac{\Lambda c^2}{3} + \frac{8\pi G}{3}\rho_0(1+z)^3}.
\end{gather}
Relation (\ref{eq3}) follows from equations (\ref{hubn5}) and (\ref{hubn6}) in the presence of DEV, and (\ref{eq4}) follows from equations (\ref{hubn8}) with zero DEV. For transformations, we should use the relation for hyperbolic functions

\begin{gather*}
    \coth(x)=\frac{\sqrt{1+\sinh^2(x)}}{\sinh(x)}.
\end{gather*}
To satisfy the condition of equality $H_\alpha(z_{r})=H(z_{r})$, at $z=z_{r}$ it is required that modern densities be different, specifically:

\begin{gather}
\label{eq5}
\rho_{0\alpha}(1+z_{r})^{\frac{3\alpha}{1+\alpha}}=\rho_0(1+z_{r})^3,
\qquad \rho_{0\alpha}=\rho_0(1+z_{r})^{\frac{3}{1+\alpha}}.
\end{gather}
This condition is used here to fulfill the equality of $H$ at the time of recombination, instead of introducing the DEV correction of the solution in the previous consideration.

By identifying the computed modern $H$ values with the observed values,

\begin{gather}
\label{eq6}
H_{0\alpha} = H_{0}^{Loc},\qquad 
H_0 = H_{0}^{Dist}, 
\end{gather}
we obtain, using (\ref{eq3})–(\ref{eq5}), an equation for determining the $\alpha$ value that eliminates the HT in the form

\begin{gather}
\label{eq7}
      \frac{(H_{0}^{Loc})^2 - \frac{\Lambda c^2}{3}}{(H_{0}^{Dist})^2 - \frac{\Lambda c^2}{3}} = \frac{\rho_{0\alpha}}{\rho_0}= (1+z_{r})^{\frac{3}{1+\alpha}}
\end{gather}
The solution of this equation determines the value of the desired parameter $\alpha_{HT} \approx 24$, which coincides with the value of this parameter from (\ref{ht1}). In fact, equations (\ref{ht1}) and (\ref{eq7}) for determining the $\alpha_{HT}$ value are identical. For their identity, it is necessary and sufficient that the following equalities hold true

\begin{gather}
\label{eq7a}
\sinh^2{\left[\coth^{-1}{\left(\frac{H_{0}^{Loc}}{\sqrt{\Lambda c^2/3}}\right)} \right]}= \frac{1}{\left(\frac{H_{0}^{Loc}}{\sqrt{\Lambda c^2/3}}\right)^2-1}, \\
\sinh^2{\left[\coth^{-1}{\left(\frac{H_{0}^{Dist}}{\sqrt{\Lambda c^2/3}}\right)} \right]}= \frac{1}{\left(\frac{H_{0}^{Dist}}{\sqrt{\Lambda c^2/3}}\right)^2-1}.
\end{gather}
It can be easily seen that both equalities hold if the following relation holds true for inverse hyperbolic functions

\begin{equation}
  \label{eq7b}
\sinh{(\coth^{-1}x)}  = \frac{1}{\sqrt{x^2-1}},\quad  
\coth^{-1}x = \sinh^{-1}\frac{1}{\sqrt{x^2-1}}.
\end{equation}
To prove the validity of this equality, we use the representation of inverse hyperbolic functions in terms of logarithms in the form \cite{rg}

\begin{equation}
   \label{eq7с}
\coth^{-1}x=\frac{1}{2}\left[\ln\left(1+\frac{1}{x}\right)
-\ln\left(1-\frac{1}{x}\right)\right],\quad \sinh^{-1}y=\ln(y+\sqrt{y^2+1}).
\end{equation}
From here, as $y=\frac{1}{\sqrt{x^2-1}}$ follows equality (\ref{eq7b}) and
the identity of equations (\ref{ht1}) and (\ref{eq7}).

\subsection{\texorpdfstring{$\alpha(z) = \frac{\alpha}{(1+z)^{\gamma}}, \quad \Lambda \neq 0$}{}}

Let us consider a model with a redshift-dependent relationship between the densities $\rho_m$ and $\rho_{DEV}$. We assume the parameter $\gamma \ll 1$ so that the values of the $\alpha(z)$ function do not differ much in the era of recombination and at the present time. Then, instead of the second equation in (\ref{eq1}), with account of (\ref{hubn5}), we obtain

\begin{equation}
   \label{eq8}
 \frac{\dot\rho}{\rho}=-\frac{3\alpha(z)}{1+\alpha(z)}\frac{\dot a}{a}
 =\frac{3\alpha}{\alpha+(1+z)^\gamma}\,\frac{\dot z}{1+z}.
\end{equation}
The solution to this equation has the form

\begin{equation}
   \label{eq9}
 \frac{\rho_\alpha}{\rho_{0\alpha}}= \left[\frac{1+\alpha}{1+\alpha(1+z)^{-\gamma}}\right]^{3/\gamma}.
\end{equation}
In the absence of DEV, $\alpha\rightarrow\infty$, and the formula for the density variation in a flat dusty universe takes the form
\begin{equation}
   \label{eq10}
 \frac{\rho}{\rho_{0}}=(1+z)^3.
\end{equation}
To obtain (\ref{eq9}), the analytical value of the indefinite integral was used \cite{rg}

\begin{equation}
   \label{eq11}
3\alpha \int\frac{dz}{(1+z)[\alpha+(1+z)^\gamma]} 
=-\frac{3}{\gamma}\ln(1+\alpha(1+z)^{-\gamma}).
\end{equation}
In this case, the integration constant was chosen in such a way that, for $\gamma=0$, the formulas were reduced to those obtained earlier. From the first relation in (\ref{eq1}), we obtain the following expressions for the Hubble parameter:

\begin{gather}
\label{eq12}
    H_{\alpha}(z) = \sqrt{ \frac{\Lambda c^2}{3} + \frac{8\pi G}{3}\rho_{0\alpha} \left[\frac{1+\alpha}{1+\alpha(1+z)^{-\gamma}}\right]^{3/\gamma}},\quad
   H(z) = \sqrt{\frac{\Lambda c^2}{3} + \frac{8\pi G}{3}\rho_0(1+z)^3}.  
\end{gather}
To fulfill the boundary condition at the level of recombination $H_{r\alpha} = H_r$, different densities in the modern era are required

\begin{gather}
\label{eq13}
    \rho_{0\alpha} \left[ \frac{1+\alpha }{1+\alpha (1+z_{r})^{-\gamma}} \right]^{\frac{3}{\gamma}} = \rho_0 (1+z_r)^3.
\end{gather}
The values of the Hubble constant in both models can then be written as

\begin{gather}
\label{eq14}
    H_{\alpha}(z) = \sqrt{ \frac{\Lambda c^2}{3} + \frac{8\pi G}{3}\rho_{0} \left[\frac{1+\alpha(1+z_r)^{-\gamma}}{1+\alpha(1+z)^{-\gamma}}\right]^{3/\gamma}(1+z_r)^3},\quad  H(z) = \sqrt{\frac{\Lambda c^2}{3} + \frac{8\pi G}{3}\rho_0(1+z)^3}. 
\end{gather}
For $z=0$, using conditions (\ref{eq6}), we obtain from (\ref{eq14}) an equation for finding the $\alpha_{HT}$ value for different values of $\gamma$:

\begin{equation}
\label{eq15}
    \frac{(H_{0}^{Loc})^2 - \frac{\Lambda c^2}{3}}{(H_{0}^{Dist})^2 - \frac{\Lambda c^2}{3}} \; \left[\frac{1+\alpha }{1+\alpha (1+z_{r})^{-\gamma}} \right]^{\frac{3}{\gamma}} = (1+z_r)^3.
\end{equation}
From this equation, for specific $\gamma$ values, we numerically find the values of the desired coefficient $\alpha_{HT}$ that eliminate the HT:
\begin{equation*}
    \begin{aligned}
        &\gamma = 0,235  \quad  &&\alpha_{HT} \approx 60,9\\
        &\gamma = 0,1  \quad  &&\alpha_{HT} \approx 34,5\\
        &\gamma = 0,01  \quad  &&\alpha_{HT} \approx 24,8\\
        &\gamma \to 0  \quad  &&\alpha_{HT} \to 23,9
    \end{aligned}
\end{equation*}
The larger the parameter $\gamma$, the faster the $\alpha(z)$ function decreases with increasing redshift, increasing the contribution of the additional component DEV in the early stages of the evolution of the Universe. At the same time, the current $\alpha_{HT}$ value and the contribution to DE from DEV that eliminates the HT are decreasing.

The age of the Universe $t_0$ where there is no HT depends on the choice of the pair ($\gamma,\;\;\alpha_{HT}$), where only the parameter $\gamma$ is independent. Only at one $\gamma$ value
this age coincides with the age from Table 1 obtained on the basis of observational data \cite{age}. To calculate the age in the model, the following relationships discussed above are used:

\begin{gather}
\label{eq16а}
    z+1=\frac{a_0}{a}=\frac{1}{x}, \quad x=\frac{a}{a_0},\quad \rho_{0\alpha}=\frac{\rho_m}{0.3},\quad 
 H=\frac{\dot a}{a}, \quad t_0=\int_0^{a_0}\frac{da}{a\,H}=
 \int_0^{1}\frac{dx}{x\,H_{\alpha}(x)},\nonumber \\
  H_{\alpha}(x) = \sqrt{ \frac{\Lambda c^2}{3} + \frac{8\pi G}{3}\rho_{0\alpha} \left[\frac{1+\alpha}{1+\alpha\,x^{\gamma}}\right]^{3/\gamma}}
\end{gather}
Calculating this integral numerically, we find that the calculated age coincides with the observed age $t_0$ from Table 1 for $\gamma \approx 0.235$ and $\alpha_{HT} \approx 61$.

\section{Conclusions}

We have considered a cosmological model in which the DE contribution consists of two components: Einstein’s cosmological constant $\Lambda$ and a small variable component $\Lambda_{V}$ (DEV) associated with the scalar field as a remnant from the inflationary stage in the early Universe. It was assumed that the energy of the variable component is uniquely related to the density of matter formed from the scalar field in the early stages of the evolution of the Universe. For simplicity, we considered an extension of the standard $\Lambda CDM$ model with a linear relationship between energy densities in the form $\rho_{m} = \alpha(z) \rho_{DEV}$, where the proportionality coefficient, in the general case, depends on the redshift. This model does not have the HT problem, which is presumably the result of inaccurate extrapolation in the calculation of the current value of the Hubble constant from observations of CMB fluctuations. We have solved the Friedmann equation under the assumption that there is an additional component of DE and its connection with matter, and we have considered several $\alpha(z)$ functions under which the HT problem is eliminated. The current DEV energy density required to explain the HT phenomenon is small with respect to the cosmological constant $\Lambda$, so it affects the expansion of the Universe only at large $z$, when the contribution of Einstein’s constant decreases.

At present, the situation is opposite, $\Lambda \gg \Lambda_{V}$, because the decrease in the density of matter in the process of cosmological expansion determines the transition from the quasi-Friedmann stage of expansion to the quasi-Sitter stage. In the case of a constant $\alpha_{HT}$ coefficient, the contradiction is eliminated at $\alpha_{HT}\approx 24$. At the same time, the density $\rho_{DEV}$ at the present epoch corresponds to the relative contribution $\Omega_{DEV} \approx 0.014$. The best agreement between the present age of the Universe and the model value occurs with a weak increase in the contribution $\rho_{DEV}$ with redshift, and a large present value of $\alpha_{HT}\approx 61$, which corresponds to a smaller present contribution to the density $\Omega_{DEV} \approx 0.0055$.

The current values of the $\Lambda CDM$ model parameters were determined from measurements of the CMB fluctuations by the WMAP and Planck experiments. This procedure is very complicated and is based on finding extrema in a multi-parameter space. Our model, which eliminates the HT problem, if it actually
exists, involves some changes in the procedure for finding the cosmological parameters, so the cosmological parameters obtained in this case may change slightly.

In our model, DM must be represented by a wide range of particle masses, in contrast to the CDM model with particles of the same mass, which is usually considered. Assuming that thermodynamic equilibrium existed in the early stages of the expansion, the number of light DM particles should be close to the number of CMB photons. Knowing the relative contribution to the CMB energy, we can roughly estimate
that the smallest mass of DM particles should not currently exceed $~ \left(\Omega_{DEV}/\Omega_{CMB} \right)^{1/4} \times kT_{CMB} \sim 5\cdot 10^{-4}\;$ eV.

\section*{Acknowledgments}

The authors are grateful to O.Yu. Tsupko for useful discussions.

\section*{Funding}

GSBK’s study was in part supported by the Russian Foundation for Basic Research, grant no. 20-52-12053.

\printbibliography

\end{document}